\documentclass[useAMS,fleqn,usenatbib]{mnras}

\usepackage{newtxtext,newtxmath}

\usepackage[T1]{fontenc}
\usepackage{ae,aecompl}

\usepackage{graphicx}	

\newcommand{\be}{\begin{equation}}
\newcommand{\ee}{\end{equation}}
\newcommand{\bea}{\begin{eqnarray*}}
\newcommand{\eea}{\end{eqnarray*}}

\title[Discovery and description of two young open clusters]{Discovery and description of two young open clusters in the primordial group of NGC 6871 }

\author[Casado and Hendy]{
Juan Casado$^{a}$\thanks{E-mail: juan.casado@uab.cat, yasserhendy@nriag.sci.eg},
Yasser Hendy$^b$
\\
$^a$Facultad de Ciencias, Universitat Autònoma de Barcelona (UAB), 08193, Bellaterra, Barcelona, Spain\\
$^b$Astronomy Department, National Research Institute of Astronomy and Geophysics, (NRIAG), 11421 Helwan, Cairo, Egypt\\
}

\date{Accepted XXX. Received YYY; in original form ZZZ}

\pubyear{2022}

\begin{document}
\label{firstpage}
\pagerange{\pageref{firstpage}--\pageref{lastpage}}
\maketitle

\begin{abstract}
A primordial group of open clusters containing NGC 6871 is confirmed and described through
Gaia DR3 data and the previous literature. It is a star-forming complex containing at least six
young OCs, including Teutsch 8, FSR 198 and Biurakan 2. Two nearby OCs (Casado 82 and
Casado-Hendy 1) are newly identified and studied in detail and found to be also members of the
cited group. The parameters of the components are sufficiently similar to postulate the case of at
least six clusters born from a single GMC. None of the cluster pairs of the group seems to be an
authentic binary cluster, with the possible exception of the candidate pair Teutsch 8/FSR 198.
Instead, NGC 6871 seems to be disintegrating, and the primordial group members appear to be
dispersing out rapidly. Searching for new open clusters in the vicinity of young or grouped OCs
using Gaia data is an efficient strategy to find new associated OCs forming primordial groups.
\end{abstract}



\section{Introduction}
\label{sec:intro}
Stars form on a wide range of scales, giving rise to gravitationally bound clusters and loose
aggregates that form hierarchical structures (Elmegreen et al. 1999). Open clusters (OCs) originate
from the gravitational collapse of gas and dust within giant molecular clouds (GMC). At least
some of them are born in groups (Bica et al. 2003; Camargo et al. 2016; Casado 2021), forming
the so-called primordial groups. For example, Chupina \& Vereshchagin (2000) studied the star-
forming region in the vicinity of the Orion nebula and found that the OCs Trapezium, NGC 1997
and NGC 1980, among other star clumps, have parallel PMs. Primordial groups seem to disperse
across the Galactic disc in a relatively short time (de La Fuente Marcos \& de La Fuente Marcos
2009, 2010; Grasha et al. 2015). Studies of grouping among OCs provide clues to understanding
hierarchical star formation in the Galactic disc and the subsequent dynamical evolution of OCs.
Moreover, they allow us to compare our Galaxy with nearby galaxies, such as the Magellanic
Clouds (Efremov \& Elmegreen 1998), concerning these subjects.\\

The ESA's Gaia mission has started a new era of precision astrometry that allows, among others,
the study of galactic clusters with unprecedented accuracy and the discovery of new OCs (Gaia
Collaboration, 2020; Cantat-Gaudin et al., 2020). The complete Gaia Data Release 3 (Gaia DR3)
was published on 13 June 2022. Gaia DR3 provides the position and apparent magnitude for $\sim1.8$
billion sources and the parallax (plx), proper motion (PMs), and BP-RP colour for $\sim1.5$ billion
sources \footnote{\url{https://vizier.cds.unistra.fr/viz-bin/VizieR?-source=I/355/}}. The improved precision
in plx and, especially, in PMs vs. Gaia DR2 offers the opportunity to revisit the OC population and
improve their characterization. Yalyalieva et al. (2020) studied the star formation history of the Sco
OB1 association using Gaia DR2 and identified some related young clusters, including NGC 6231
and Trumpler 24. Casado (2022) investigated the youngest OCs (age < 0.01 Gyr) in Tarricq et al.
(2021) and obtained that $\sim71\%$ of them remain in their primordial groups. A similar study of older
OCs (age > 4 Gyr) shows they are essentially alone. These and other results in the same work lead
to the Primordial Group hypothesis: only young OCs can be multiple, and old OCs are virtually
alone, since the gravitational interaction between OCs in primordial groups is extremely weak, and
the probability of gravitational capture of two OCs without disruption or merger is very low. In
previous works, we observed that searching for associated clusters around a given group (Casado
2021) or very young stellar clusters (Casado 2022), is an effective method for discovering new
associated OCs.\\

NGC 6871 is a large, rich OC with some substructure and at least two cores (Fig. 1). De la Fuente
Marcos \& de la Fuente Marcos (2009) and Conrad et al. (2017) proposed that NGC 6871 forms a
candidate binary cluster with Biurakan 1, but modern astrometric measurements of the Gaia
mission show that this assumption was flawed. Liu \& Pang (2019) proposed NGC 6871 and
Gulliver 17 as candidate members of an OC pair. However, most astrometric parameters of both
OCs, particularly the PMs, are discordant, too.\\

On the other hand, there is recent evidence of a primordial group comprising NGC 6871. The
cluster pair FSR 198/Teutsch 8 can be found in the works of Piecka \& Paunzen (2021) and Casado
(2022). They are likely members of a triple primordial group comprising FSR 198, Teutsch 8, and
NGC 6871 (Casado 2022). Biurakan 2 has also been considered a candidate member of the group
(Piecka \& Paunzen 2021).\\

The initial objective of the present work was to revisit the primordial group of NGC 6871 in order
to take advantage of Gaia DR3 results to obtain a better characterization of this star formation
complex and the individual component clusters. While doing so, we have identified two new OCs
(Casado 82 and Casado-Hendy 1) that we characterize in detail. The outline is as follows. In
section 2, we summarize the methodology employed. In Section 3, we revisit NGC 6871. The
main characteristics of its primordial group are summarized in Section 4. Sections 5 and 6 are
devoted to a detailed description of the two new clusters of that group. Section 7 compares NGC
6871 and the two new clusters to determine the relationships between them, and Section 8
highlights the main conclusions.\\

\begin{figure*}
\includegraphics[height=6cm]{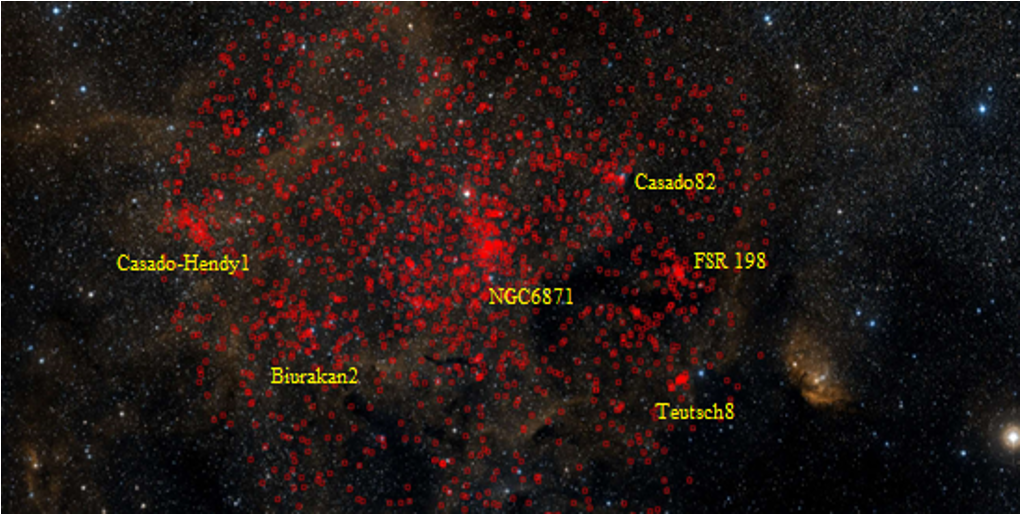}
\caption{Some likely members of the HC primordial group. Constraints plx 0.41 to 0.61 mas,
PMRA -2.8 to -3.8 mas/yr, PMDec -5.9 to -7.0 mas/yr. G<18.}
\label{figure_1}
\end{figure*}

\section{Methodology}
\label{sec:Methodology}
The methods applied to select and study candidate OC pairs have been detailed previously (Casado
2021, 2022). However, we recapitulate here the general methodology. We start with a candidate
member of a hypothetical group of Galactic OCs from the literature. For each of these candidates,
we look for close correlations between coordinates, PMs and plx for all OCs within the studied
area. For example, if two OCs are at a distance of less than 100 pc (Conrad et al. 2017; Soubiran et
al. 2019; Liu \& Pang 2019; Casado 2021), the rest of their astrometric data are compared. If there
is an overlap of the data, considering uncertainty intervals of less than $3\sigma$, both OCs are considered
candidate members of the same group. Those candidates are characterized using the most accurate
parameters of the reports using Gaia data. When existing data are incomplete or inconsistent
between different authors, the OCs are re-examined using Gaia DR3 to obtain a new set of the
corresponding parameters.\\

Perhaps the most direct way to search for possible OCs linked to each studied OC is to plot a graph
of the Gaia sources that satisfy the examined OC constraints for the studied field. In this way, we
can obtain plots similar to Fig. 1, showing (or not) any related OCs. These plots are free of most of
the noise from the unrelated field stars and, in the present case, revealed the two new OCs
described in this work: Casado 82 and Casado-Hendy 1.\\

Concerning the individual OC parameterization, we downloaded the selected data from Gaia DR3
for NGC 6871 and other OCs of its primordial group. To use only the best quality sources, we
removed the stars with RUWE > 1.4 and with G > 18, as the plx and PM errors increase
exponentially with magnitude. To obtain the likely members of each OC, we used the pyUPMASK
algorithm, which performed very well for contaminated clusters, outperforming UPMASK (Krone-
Martins \& Moitinho 2014), as shown in Pera et al. (2021). To determine the fundamental
parameters (age, distance, extinction, and metallicity), we used the ASteCA algorithm (Perren et
al. 2015, ASteCA website\footnote{\url{http://asteca.github.io/}}). We selected the most probable members for
each cluster, i.e. stars with a membership probability (p) larger than 0.7.\\

The radial density profile (RDP) is determined by counting stars in concentric rings around the
centre of clusters. We calculated the density of each sector by dividing the number of stars in the
ring by its area (Tadross and Hendy 2016; Hendy and Bisht 2021; Tadross and Hendy 2022). The
density profile usually represents an approximately exponential decrease of the surface density of
the cluster from its centre outwards and the stability of the background density beyond the cluster
boundary. The limiting radius $\it r_{lim}$ is the cluster radius determined visually as the radius where the
RDP reaches the stability of the background density.\\

To convert the metallicity $[M/H]$ to abundance $z$. we used the analytic equations of Bovy\footnote{\url{https://github.com/jobovy/isodist/blob/master/isodist/Isochrone.py/}} for PARSEC isochrones (Bressan et al. 2012). The equations are expressed as:
\begin{equation}
z_{\rm x}={10^{{\rm [M/H]}+\log \left(\frac{z_{\odot}}{0.752 - 2.78 \, z_{\odot}}\right)}}
\end{equation}      
\begin{equation}
z=\frac{(z_{\rm x}-0.2485 \, z_{\rm x})}{(2.78 \, z_{\rm x}+1)}.
\end{equation} 

Here, $z$  refers to the elements heavier than helium, $z_x$ is the intermediate operation function, and $z_{\odot}$ is the solar metallicity, which was adopted as 0.0152.\\

We used the pyUPMASK algorithm to determine member stars. The intrinsic distance modulus of the cluster $(m-M)_0$ from the ASteCA code, is used to convert the G-magnitudes of the member stars to absolute magnitudes. To obtain the mass of every member star, we used the high-degree polynomial equation between the absolute magnitudes and the masses of the main sequence stars from the isochrones of the same age and metallicity (Hendy and Tadross 2021; Tadross and Hendy 2021; Tadross and Hendy 2022).

\section{Revisiting NGC 6871}
\label{sec:Revisiting NGC 6871}

Table 1 shows the main statistics of the astrometric parameters for 430 likely members of the open
cluster NGC 6871 that have been found in this study using Gaia DR3. They agree with those from
the literature (Table 2).\\

The nine reported Gaia parallaxes for NGC 6871, after offset correction of 0.029 mas for Gaia
DR2 measurements (Lindegren et al. 2018) and 0.017 mas for Gaia EDR3 measurements
(Lindegren et al. 2021), range from 0.532 mas (Poggio et al. 2021) to 0.553 mas (Hunt and Reffert
2021), with a median value of 0.543 mas (Cantat-Gaudin et al. 2020). This last value implies a
derived distance of 1.84 kpc. Our study of this OC revealed a corrected median plx of 0.53 mas
and, therefore, a distance of 1.88(8) kpc.\\

\begin{table}
\caption{Statistics of the astrometric parameters for 430 probable members of the open cluster NGC 6871, as found in this study using Gaia DR3.}
\label{tab:table_1}
\begin{tabular}{lccccc}
\hline
Parameter&          Mean   &      Median &    SD   &     Minimum&    Maximum \\
\hline             
RA(J2016)&                 301.58 &      301.55 &    0.21 &     300.95 &     302.13 \\
dec(J2016)&                35.77  &      35.77  &    0.19 &     35.30  &     36.28 \\
l &                 72.67  &      72.65  &    0.16 &     72.23  &     73.12 \\
b &                 1.99   &      2.02   &    0.19 &     1.51   &     2.54  \\
$\it plx$&          0.51   &      0.51   &    0.03 &     0.41   &     0.60  \\
$\mu_{\alpha}$&     -3.09  &      -3.09  &    0.10 &     -3.40  &     -2.80 \\
$\mu_{\delta}$ &    -6.43  &      -6.44  &    0.17 &     -6.93  &     -5.96 \\
\hline
\end{tabular}
\break
\end{table}

There is consensus that NGC 6871 is very young, to the point that there are visible rests of the
GMC from which it was born. All reported ages range from 5.5 Myr (Cantat-Gaudin et al. 2020)
to 11.6 Myr (Dias et al. 2021), although eight out of the twelve reported ages go from 8.9 Myr
(Loktin and Matkin 1994) to 10 Myr (Dambis et al. 1999; Karchenko et al. 2013).\\

Revisiting the spectrophotometric properties of NGC 6871 using Gaia DR3, we also obtained
results similar to previous literature. The isochrone fitting of the 430 most probable members (Fig.
2) led to $log (t/yr)=6.89(2)$, i.e., an age range between 7.4 and 8.1 Myr. The estimated absolute
metal abundance z is 0.018(5). Liu \& Pang (2019) reported a z vs the sun of 0.25. However, other
authors (e.g. Gozha et al. 2012) have reported metallicities vs the sun as low as -0.33, i.e. z =
0.007, and there is no consensus on this issue.\\

The obtained extinction is $A_v=1.5$, and the distance modulus is 11.35(2) mag, pointing to a
photometric distance of 1.86(2) kpc. This value is well-matched with the maximum reported
photometric distance up to now (1.84 kpc: Cantat-Gaudin et al. 2018), and in excellent agreement
with the above-stated parallax distances, including our own calculation.\\

Although NGC 6871 is a well-studied cluster (207 references found), there is no consensus on its
RV. Reported RVs for the ensemble of this OC range from -10.5 km/s (Kharchenko et al. 2013) to
14.6 km/s (Tarricq et al. 2021). The literature search for individual stars of NGC 6871 led to
comparable results. Mermilliod's (1984) compilation shows that most of the historical
measurements of stars in the OC zone are between -26 and 5.6 km/s. Huang et al. (2010) measured
the RV of 25 member stars on two different nights. 17 RVs of the first night ranged from -20 to -4
km/s, and 16 RVs of the second night ranged from -29 to -17 km/s.\\

Therefore, we have searched the RV data on the star members of NGC 6871 using Gaia DR3 and
found that most (24) of the individual RV measurements range from -28 to 20 km/s. The high
dispersion on the RV of the individual stars makes it difficult to obtain an accurate value for the
whole cluster and suggests that the young system is in a phase of violent relaxation. Anyway, the
median of our RVs is -4.4 km/s, with a median absolute deviation of 19 km/s. These statistics are
compatible with the previous reports.\\

The maximum distance from a cluster member ($r_{max}$) to the mean position of NGC 6871 is 34 arcmin (18 pc). The limiting radius, including the cluster corona, was estimated to be ($r_{max}=65$ arcmin (35 pc) from Fig. 3. We estimated that the total mass ($\sum\,mass$), average star mass ($\sum\,mass /N$, $N$ is number of members), minimum star mass, and maximum star mass are $832\,M_{\odot}$, $1.94\,M_{\odot}$, $0.8\,M_{\odot}$ and $18\,M_{\odot}$, respectively, as detailed in Section 2. The average star mass is a biased maximum value, as NGC 6871 has more stars than the 430 identified members, but can be used for comparative purposes.\\

\begin{figure}
\includegraphics[height=9cm]{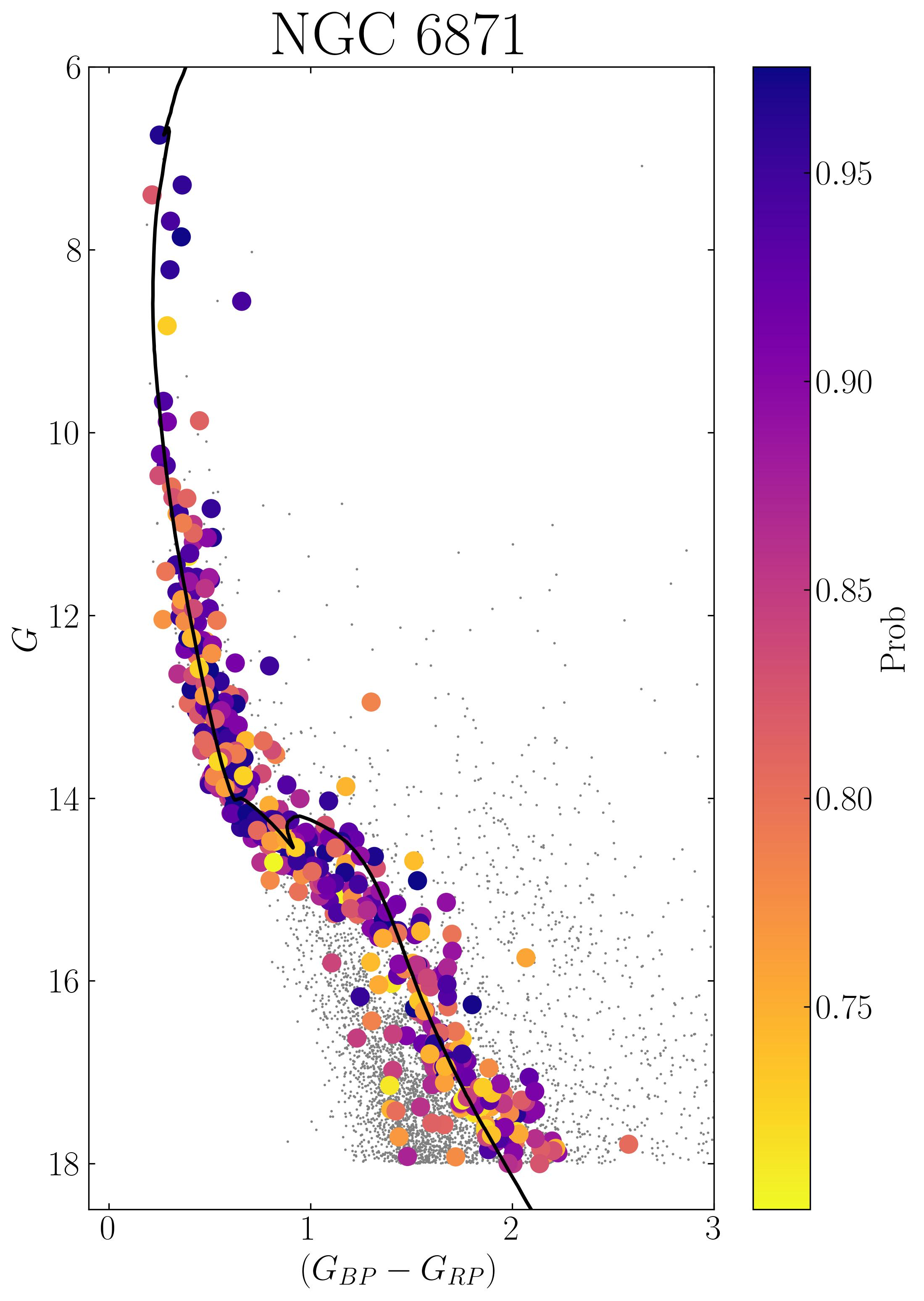}
\caption{The CMD and the best fitting isochrone of NGC 6871 down to G = 18, using ASteCA (Perren et
al. 2015). Colours of the symbols indicate their membership probabilities (right column). Grey dots are for
the field stars.}
\label{figure_2}
\end{figure}

\begin{figure}
\includegraphics[height=6cm]{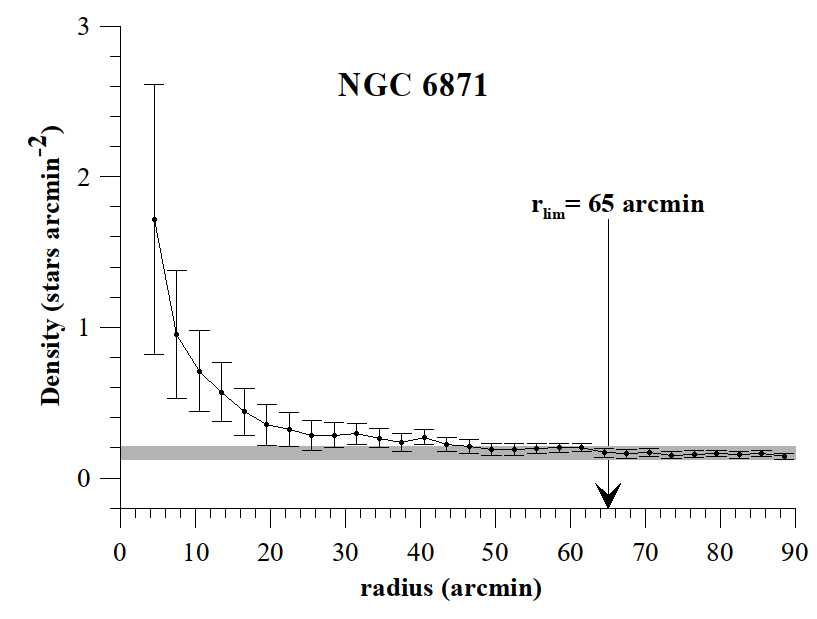}
\caption{The RDP of the star field surrounding NGC 6871. The error bars represent the $1⁄\sqrt N$ error,
while the horizontal grey line denotes the background field density. The black arrow indicates the limiting
radius $r_{lim}$.}
\label{figure_3}
\end{figure}

\section{The primordial group of NGC 6871}
\label{sec:The primordial group of NGC 6871}
NGC 6871 dominates a primordial group as the most populated
cluster. The group appears to be formed by at least six OCs (Table
2). NGC 6871 can be associated with the embedded cluster Teutsch 8, despite minor
differences in the reported distances and PMs. For NGC 6871, the
reported photometric distances range from 1.51 kpc (Glushkova
et al. 1997) to 1.84 kpc (Cantat-Gaudin et al. 2018), while for
Teutsch 8, literature distances span from 1.60 kpc (Dias et al. 2002)
to 1.98 kpc (Cantat-Gaudin et al. 2020). Their similar parallaxes
suggest that both OCs are likely to be at a compatible distance
rom the ensemble of data ($\sim 1.9$ $\it kpc$). There is consensus on
the mean PMs derived from Gaia data, which seem coherent for
both OCs. Quantitatively, $\triangle$PM/$plx$ (and $\triangle$PM d) are $\sim0.9$ $yr^{-1}$,
which implies a minor difference in tangential velocities (< 5 $km/s$).\\

\begin{table*}
\caption{Candidate member’s properties of the Primordial Group of NGC 6871. Column headings: 1. Group acronym; 2. OC name; 3. Galactic longitude; 4. Galactic latitude; 5. Parallax; 6. Photometric distance; 7. PM in right ascension; 8. PM in declination; 9 OC radius; 10. Number of member stars; 11. Age; 12. Radial velocity. Abbreviations: $^a$ radius containing 50\% of members; $^b$ maximum cluster member's distance to the average position; $^c$ see text; + et al.}
\label{tab:table_2}
\begin{tabular}{llccccccccccl}
\hline
1  &   2                &  3       & 4       &   5       &   6     & 7                & 8                & 9        &  10    &  11   &  12      &   13                           \\ 
Gr &   $\it OC$               &  $\it l$       & $\it b$       &   $\it plx$     &   $\it d$     & $\mu_{\alpha}$   & $\mu_{\delta}$   & $\it R$        &  $\it N$     &  $\it Age$  &  $\it RV$      &   $\it References$                   \\ 
   &   $\it Name$             &  $\it degree$  & $\it degree$  &   $\it mas$     &   $\it kpc$   & $\it mas$ $yr^{-1}$    & $\it mas$ $yr^{-1}$    & $\it arcmin$   &  $\it stars$ &  $\it Myr$  &  $\it km/s$    &                                \\ 
\hline
HC &   NGC 6871         &  72.66   & 2.01    &   0.514   &   1.72  & -3.13           & -6.44           & $22^a$   &  430   &  5    &          &   Cantat-Gaudin+ 2020          \\ 
   &                    &          &         &           &   1.84  &                 &                 & $22.3^a$ &  594   &       &          &   Cantat-Gaudin \& Anders 2020 \\ 
   &                    &          &         &           &   1.66  &                 &                 &          &        &  6    &  15(15)  &   Tarricq+ 2021                \\ 
   &                    &  72.67   & 1.99    &   0.51    &   1.86  & -3.09           & -6.43           & $34^b$   &  430   &  7.8  &  -4(19)  &   This work                    \\ 
\hline
HC &   Teutsch 8        &  71.86   & 2.42    &   0.49    &   1.98  & -3.49           & -6.70           & $0.4^a$  &  28    &  4    &          &   Cantat-Gaudin+ 2020          \\ 
   &                    &          &         &   0.497   &   1.97  & -3.33           & -6.76           & $2.7^b$  &  27    &  3    &  -1(5)   &   This work                    \\ 
\hline
HC &   FSR 198          &  72.18   & 2.61    &   0.49    &   2.18  & -3.6            & -6.6            & $7.6^a$  &  82    &  5    &          &   Cantat-Gaudin+ 2020          \\ 
   &                    &          &         &           &   2.04  &                 &                 &          &        &  5.4  &  12(2)   &   Tarricq+ 2021                \\ 
   &                    &          &         &   0.494   &   1.97  & -3.5            & -6.6            & $17^b$   &  97    &  4    &  -11(12) &   This work                    \\ 
\hline
HC &   Biurakan 2       &  72.75   & 1.36    &   0.54    &   1.72  & -3.17           & -6.84           & $8.3^a$  &  47    &  9    &  -20     &   Cantat-Gaudin+ 2020          \\ 
   &                    &          &         &           &         &                 &                 &          &        &       &  $-25^c$ &                                \\ 
\hline
HC &   Casado 82        &  72.61   & 2.59    &   0.493   &   2.00  & -3.24           & -6.24           & $9^b$    &  39    &  5.8  &  -22(9)  &   This work                    \\ 
\hline
HC &   Casado-Hendy 1   &  73.31   & 1.17    &   0.49    &   1.99  & -3.19           & -6.31           & $8.1^b$  &  39    &  5.9  &  -24(18) &   This work                    \\ 
\hline
\end{tabular}
\break
\end{table*}

If NGC 6871 is associated with Teutsch 8, it would also be affiliated with FSR 198 since FSR 198
appears to form a double system with Teutsch 8: All the relevant parameters are well-matched
considering the observational error (Table 2). As there were no previous reports of RV for Teutsch
8, we have re-examined it using Gaia DR3, and we have found a very likely ($p > 0.99$) member
star (Gaia DR3 2059446855803012480) with an RV measurement. Revised parameters for
Teutsch 8 are summarized in Table 2. The parallax distance of Teutsch 8 ($1.94\pm0.13$) is compatible
with the photometric distance of all the group members. The obtained extinction was $A_v=1.9$,
comparable with most of the group siblings. We also revisited FSR 198 and found two likely
members ($p=0.98$ and 0.96) with similar RV (Gaia DR3 2058698500692882560 and
2059447852234563712, respectively). The rest of the recalculated parameters are compatible with
previous reports and with their companions, except for the extinction, $A_v= 2.7(2)$, which
corresponds to a local region of denser nebulosity. The compatible RVs of NGC 6871, Teutsch 8,
and FSR 198 (Table 2) increase the likelihood of a triple group. However, there is no literature
consensus on the RV of NGC 6871, as discussed in the previous section. The ages of the three
candidate members are again compatible with a unique (and recent) origin.\\

All reported RVs of Biurakan 2 range from -19.7 km/s (Dias et al. 2014) to -24.9 (Zhong et al.
2020), with a consensus value of -22 km/s that has been frequently quoted (Vande Putte et al.
2010; Kharchenko et al. 2013; Loktin \& Popova 2017; Conrad et al. 2017). As we will detail later,
these values are virtually coincident with our RVs for Casado 82 and Casado-Hendy 1. Biurakan 2
also shows parameters that, on the whole, seem at least marginally compatible with the rest of the
group members (Piecka \& Paunzen, 2021). Therefore, the case for a primordial sextet, dubbed HC
in Table 2, seems robust enough.\\

Other candidate clusters that could be members of the same primordial group (i.e. born from the
parent GMC of the above-discussed OCs) are IC 4996, Berkeley 87, and Dolidze 3, despite
evidence supporting this possibility is not conclusive so far.

\section{A new cluster in the group: Casado 82}
\label{sec:A new cluster in the group: Casado 82}
From the inspection of the surroundings of NGC 6871, while studying its primordial group, an
additional member was discovered: Casado 82 (Fig. 1). The astrometric statistics of this new OC
are given in Table 3. From the parallax measurements of its likely members, the median Gaia DR3
parallax was 0.493 mas, which after offset correction (Lindegren et al. 2021), leads to a derived
distance of 1.96(8) kpc.\\

\begin{table}
\caption{Statistics of the astrometric parameters for 39 likely members of the new open cluster Casado 82.}
\label{tab:table_3}
\begin{tabular}{lccccc}
\hline
Parameter&   Mean&        Median&    SD&       Minimum&    Maximum \\
\hline
RA(J2016)&                 300.92 &  300.91 &  0.06 &   300.78  &  301.05  \\  
dec(J2016)&                36.03  &  36.04  &  0.07 &   35.89   &  36.15   \\
l &                 72.61  &  72.61  &  0.06 &   72.46   &  72.73   \\
b &                 2.59   &  2.59   &  0.06 &   2.45    &  2.73    \\
$\it plx$&          0.499  &  0.493  &  0.04 &   0.42    &  0.61    \\
$\mu_{\alpha}$&     -3.24  &  -3.24  &  0.09 &   -3.44   &  -3.02   \\    
$\mu_{\delta}$ &    -6.24  &  -6.26  &  0.10 &   -6.39   &  -6.00   \\      
\hline
\end{tabular}
\break
\end{table}

The isochrone fitting of the 39 members of this cluster having $p>0.7$ led to log (t/yr) =6.76(9),
i.e., an age from 4.7 to 7.1 Myr (Fig. 4). The obtained extinction was $A_v = 1.55(3)$ and the distance
modulus 11.50(6) mag, which corresponds to a photometric distance of 2.00(8) kpc. We note that
the two derived distances are well-matched.\\

The estimated z using ASteCA was 0.015(6), close to the sun's z. We cross-matched the 39
members of Casado 82 with StarHorse's catalogue (Anders et al. 2022), and we found [M/H]
values for 31 of them. From these values, Casado 82 would have a median metallicity [M/H] =
-0.11. From this median, we calculated z = 0.012, in good agreement with the ASteCA estimate.\\

One of the members of Casado 82 (Gaia DR3 2059842542555949312, $p = 0.99$) has an RV value
of -22(9) km/s found by Gaia DR3. The ASteCA code obtained a contour map on the cluster’s coordinates using the two-dimensional
kernel density analysis (Fig. 5a). Some overdensity of stars around the cluster centre is also
noticeable in the sky map of stars of G < 18 (Fig. 5b).\\

The maximum cluster member's distance to the average position ($r_{max}$ ) of Casado 82 is 9.0 arcmin
(5.2 pc). We estimated $r_{lim}= 15$ arcmin (9 pc) from Fig. 6. Note that $r_{max}$ is smaller than $r_{lim}$ .
This discrepancy seems to arise from the presence of a slight overdensity of stars at position 301.0,
36.2 (Fig. 5 a), which is seen as a small protrusion at a radius of $\sim 12$ arcmin in Fig. 6. 
We estimated that the total mass ($\sum\,mass$), average star mass
($\sum\,mass/N$, $N$ is number of members), minimum star mass, and maximum star mass are
$84\,M_{\odot}$, $2.14\,M_{\odot}$, $0.84\,M_{\odot}$ and $18\,M_{\odot}$, respectively.

\begin{figure}
\includegraphics[height=9cm]{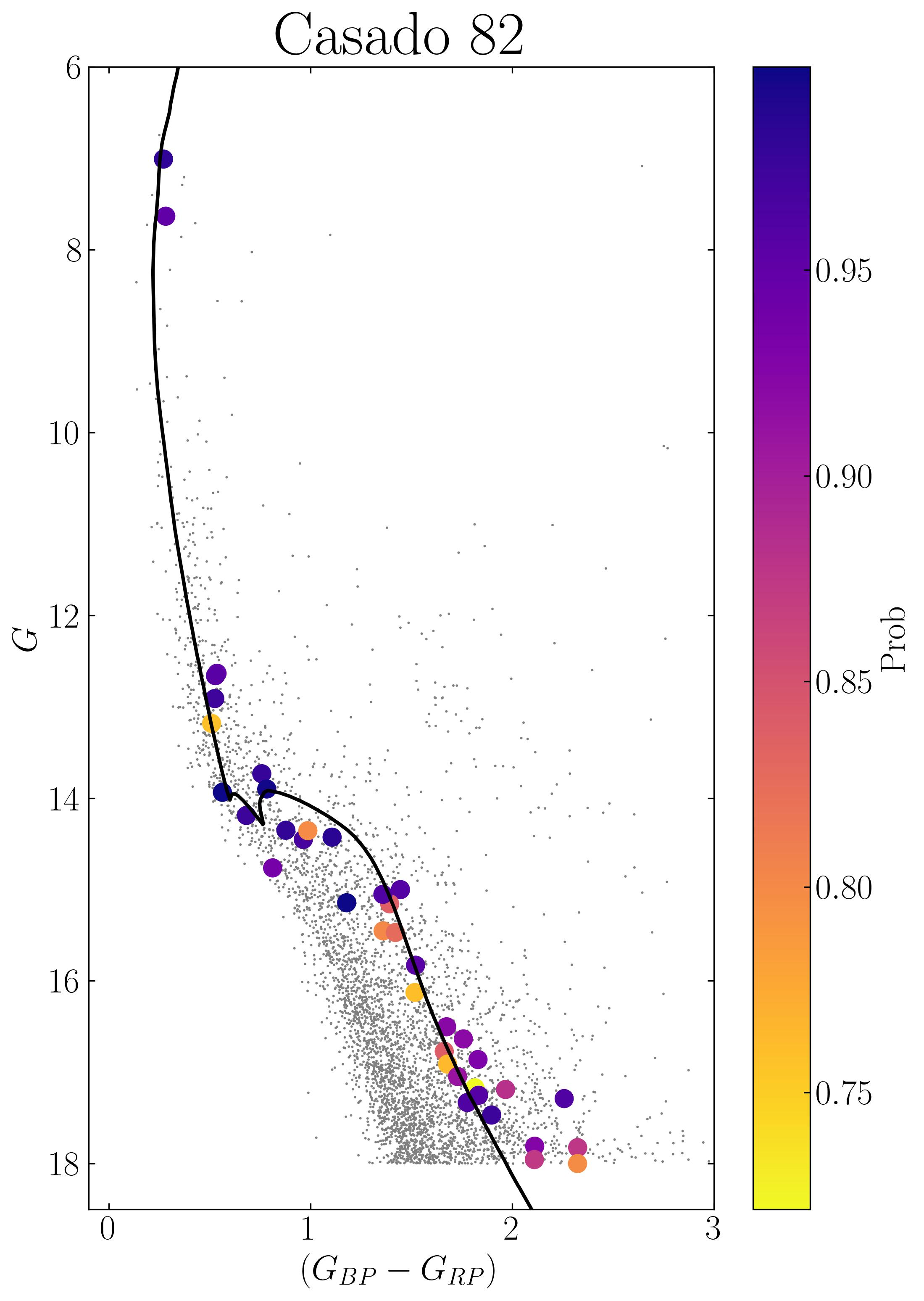}
\caption{The CMD and the best fitting isochrone of the new open cluster Casado 82 down to G = 18, using
ASteCA (Perren et al. 2015). Colours of the symbols indicate their membership probabilities (right column).
Grey dots are for the field stars.}
\label{figure_4}
\end{figure}

\begin{figure*}
\includegraphics[height=6cm]{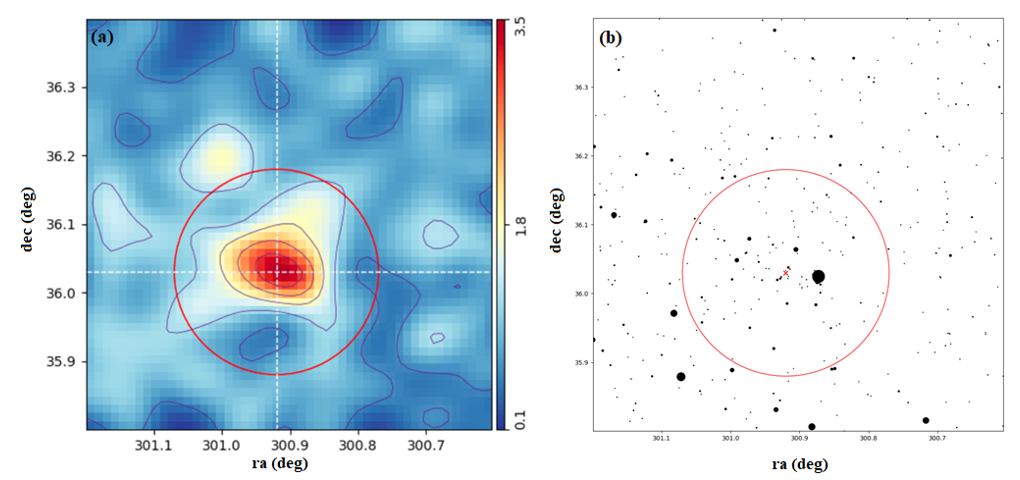}
\caption{(a): The contours map of the new open cluster Casado 82, using ASteCA algorithm. The red circle
corresponds to our $r_{max}$ estimation. (b): The sky map of Casado 82.}
\label{figure_5}
\end{figure*}

\begin{figure}
\includegraphics[height=6cm]{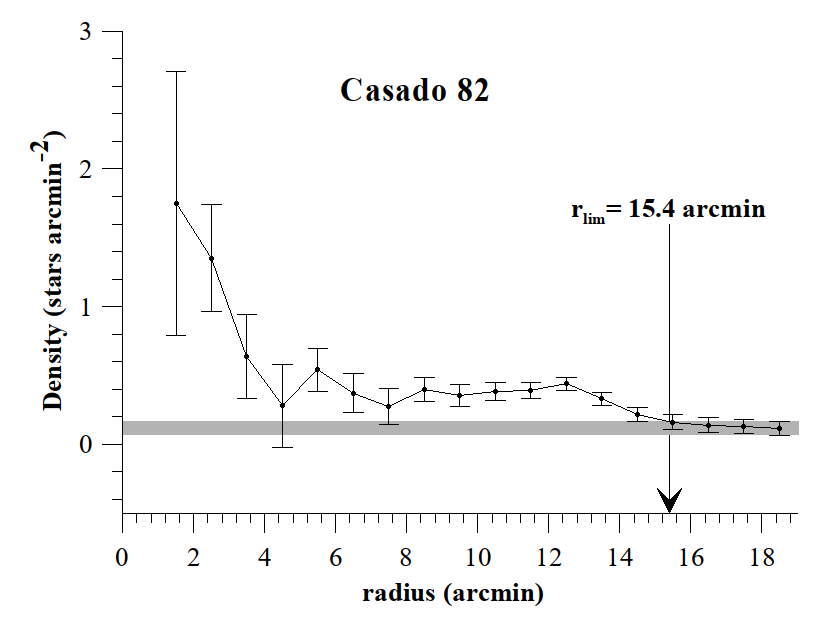}
\caption{The RDP of the star field surrounding Casado 82. The error bars represent the $1⁄\sqrt N$ error, while
the horizontal grey line denotes the background field density. The black arrow indicates the limiting radius
$r_{lim}$.}
\label{figure_6}
\end{figure}

\begin{figure}
\includegraphics[height=9cm]{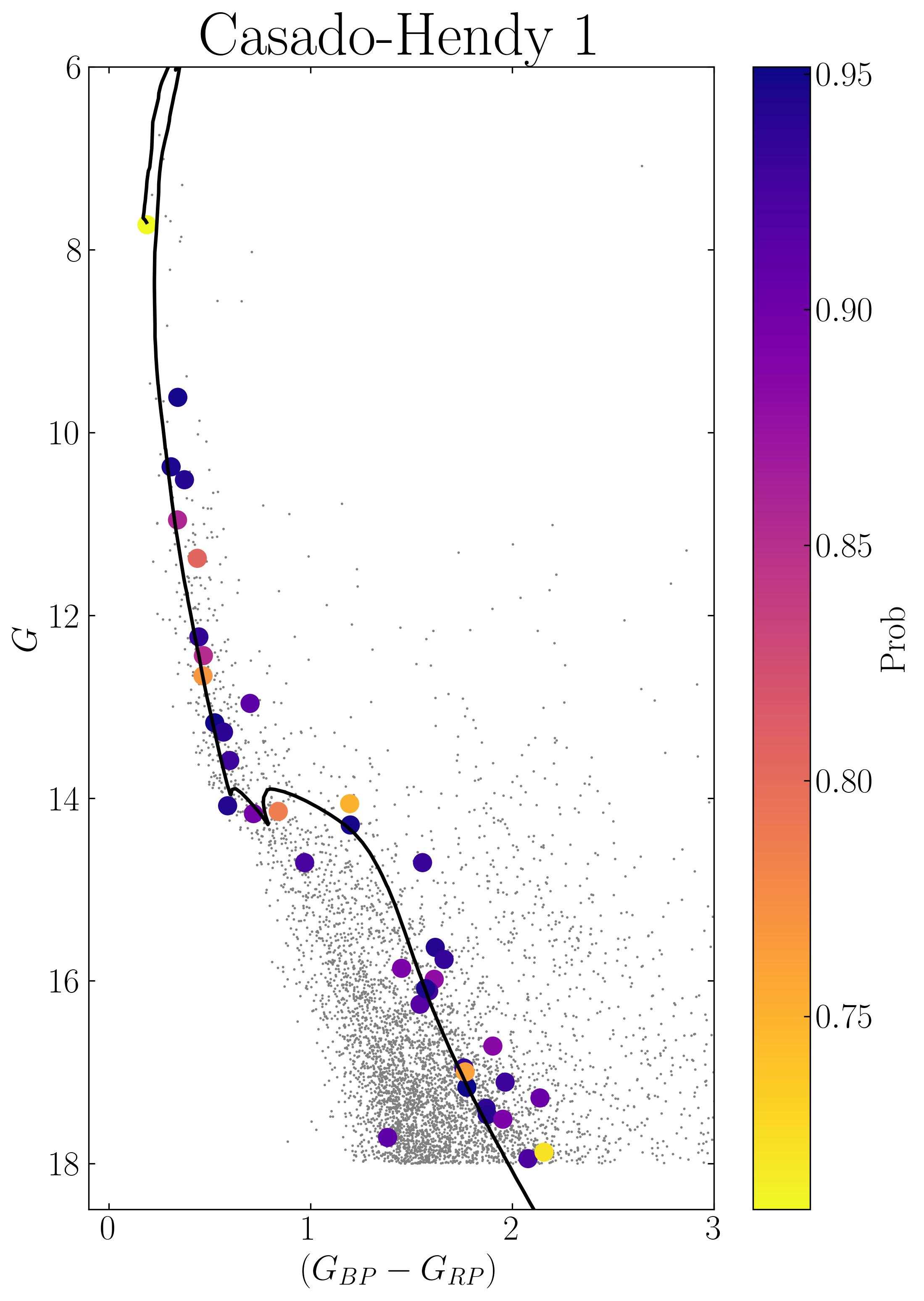}
\caption{The CMD and the best fitting isochrone of the new open cluster Casado-Hendy 1 down to G = 18,
using ASteCA (Perren et al. 2015). Colours of the symbols indicate their membership probabilities (right
column). Grey dots are for the field stars.}
\label{figure_7}
\end{figure}

\begin{figure*}
\includegraphics[height=5cm]{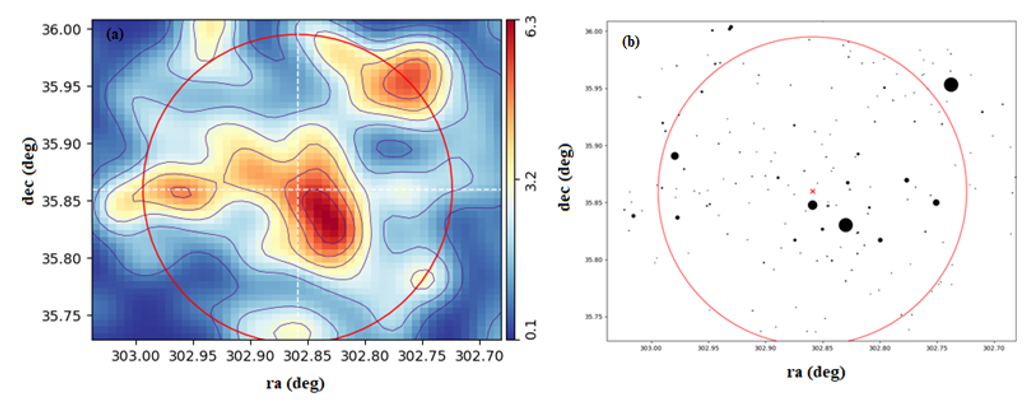}
\caption{(a): The contours map of the open cluster Casado-Hendy 1, using ASteCA algorithm. The red
circle corresponds to our $r_{max}$ estimation. (b): The sky map of Casado-Hendy 1 for G < 18.}
\label{figure_8}
\end{figure*}

\begin{figure}
\includegraphics[height=6cm]{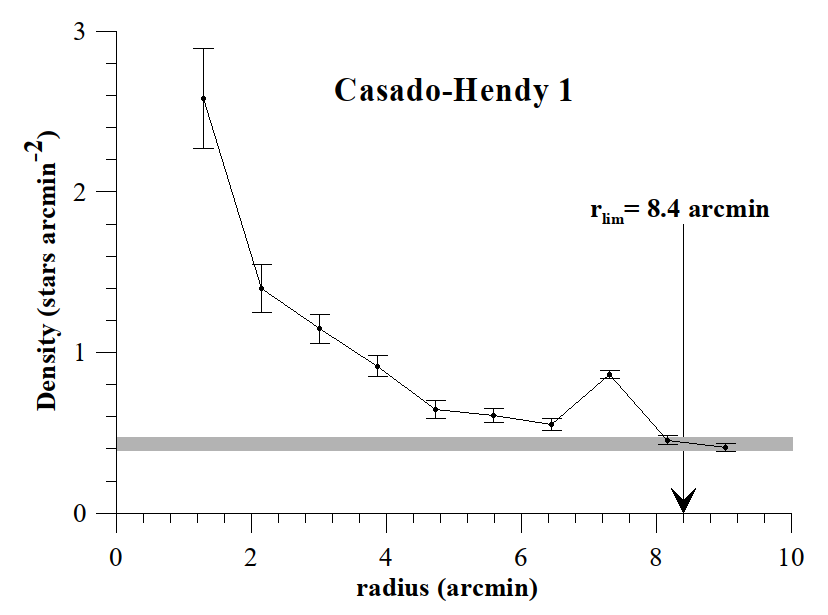}
\caption{The RDP of Casado-Hendy 1. The error bars represents the $1⁄\sqrt N$ error, while the horizontal grey
line denote the background field density. The black arrow line indicates the obtained $r_{lim}$.}
\label{figure_9}
\end{figure}

\begin{figure}
\includegraphics[height=9cm]{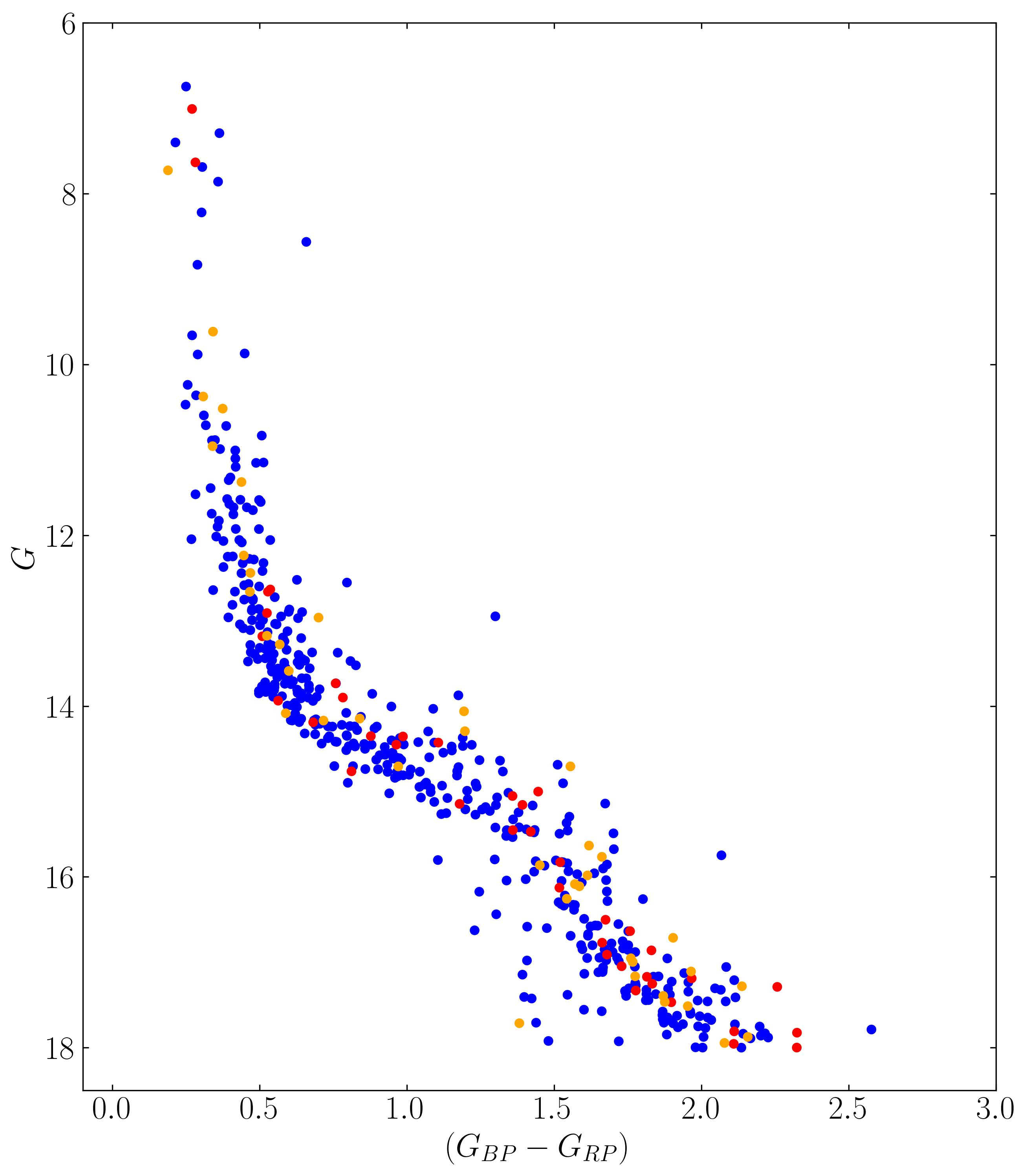}
\caption{The superposed CMDs of NGC 6871 (blue dots), Casado 82 (red dots) and Casado-Hendy 1
(orange dots).}
\label{figure_10}
\end{figure}

\begin{figure}
\includegraphics[height=7cm]{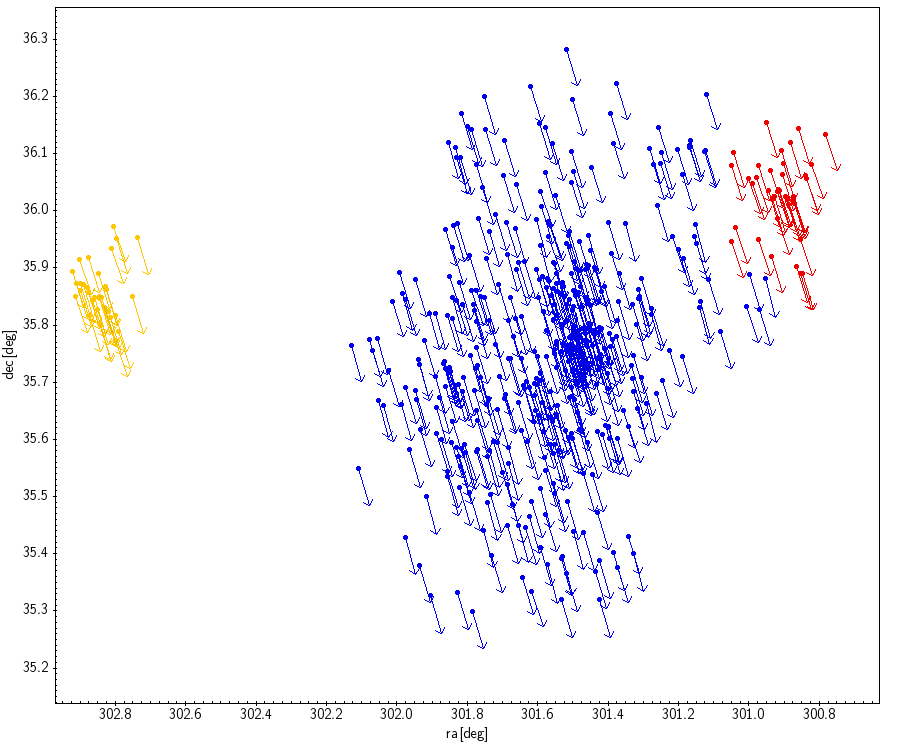}
\caption{The positions and PMs showing the comoving stars of NGC 6871 (blue arrows),
Casado 82 (red arrows) and Casado-Hendy 1 (orange arrows).}
\label{figure_11}
\end{figure}

\section{A second new cluster in the group: Casado-Hendy 1}
\label{sec:A second new cluster in the group: Casado-Hendy 1}
While plotting the NGC 6871 primordial group (Fig. 1), an additional OC was discovered: Casado-
Hendy 1, with at least 39 members having $p > 0.7$. The astrometric statistics of this new OC are
summarized in Table 4. From their parallax measurements, the median Gaia DR3 parallax was
0.49 mas, which after offset correction (Lindegren et al. 2021), leads to a derived distance of
1.97(8) kpc.\\

\begin{table}
\caption{Statistics of the astrometric parameters for 39 likely members of the new open cluster Casado-Hendy 1.}
\label{tab:table_3}
\begin{tabular}{lccccc}
\hline
Parameter&   Mean&        Median&    SD&       Minimum&    Maximum \\
\hline
RA(J2016)&                 302.86  &   302.85  &  0.06  &   302.74  &  303.02  \\  
dec(J2016)&                35.86   &   35.85   &  0.05  &   35.77   &  35.97   \\
l &                 73.31   &   73.31   &  0.05  &   73.22   &  73.42   \\
b &                 1.17    &   1.17    &  0.05  &   1.05    &  1.30    \\
$\it plx$&          0.49    &   0.49    &  0.03  &   0.43    &  0.56    \\
$\mu_{\alpha}$&     -3.19   &   -3.17   &  0.10  &   -3.43   &  -3.03   \\    
$\mu_{\delta}$ &    -6.31   &   -6.31   &  0.12  &   -6.64   &  -6.13   \\      
\hline
\end{tabular}
\break
\end{table}

The isochrone fitting of the likely members of this cluster (Fig. 7) led to $log(t/yr)=6.77\pm0.21$,
i.e., an age from 3.6 to 9.5 Myr. The obtained extinction was $A_v=1.55(3)$, and the distance
modulus was 11.49 mag, corresponding to a photometric distance of 1.99 kpc. Again, the two
derived distances are well-matched.\\

The estimated z using ASteCA was 0.017(6), close to that of the sun. We cross-matched the 39
members of Casado-Hendy 1 with StarHorse’s catalogue (Anders et al. 2022), and we found
[M/H] values for 31 of them. From these values, Casado-Hendy 1 would have a median metallicity
[M/H]= -0.11. From this median, we calculated z = 0.012, in good agreement with the ASteCA
estimate, too.\\

The ASteCA code obtained a contour map on the cluster’s coordinates using the two-dimensional
kernel density analysis (Fig. 8a). Some overdensity of stars nearby the cluster centre is also
noticeable in the sky map of stars of $G < 18$ (Fig. 8b).

The maximum cluster member's distance to the average position ($r_{max}$ ) of Casado-Hendy 1 is 8.1
arcmin (4.7 pc). We estimated $r_{lim}$ = 8.4 arcmin (4.9 pc) from Fig. 9. We note that r max is close to
$r_{lim}$ . The total mass ($\sum\,mass$), average star mass ($\sum\,mass/N$, $N$ is number of members),
minimum star mass, and maximum star mass are $93\,M_{\odot}$, $2.39\,M_{\odot}$, $0.82\,M_{\odot}$ and $16\,M_{\odot}$ ,
respectively.\\

\section{Comparing the new OCs and NGC 6871}
\label{sec:Comparing the new OCs and NGC 6871}
When we compare the CMDs of the three OCs by superposition, the obtained overlapping is
virtually a perfect match (Fig. 10), which endorses the hypothesis that both OCs were born from a
single GMC and belong to the same primordial group. The minor differences found in the
parameters derived by isochrone fitting should be -and mostly are- within the error margins of the
method.\\

We have found that metal abundances are z = 0.018, 0.015 and 0.017 for NGC 6871, Casado 82
and Casado-Hendy 1, respectively. Note the three values of metal abundance are very similar and
close to that of the sun. The abundances obtained from StarHorse’s catalogue for the two new OCs
are virtually identical (0.012). These results confirm that the three open clusters are siblings with
similar chemical compositions.\\

If we compare the PMs of the three clusters (Fig. 11), the inevitable conclusion is the same: all
their likely members share similar PMs, which reinforces the conclusion that they belong to the
primordial HC group, although they are probably not a gravitationally bound triplet, as we discuss
below.\\

\begin{figure}
\includegraphics[height=7cm]{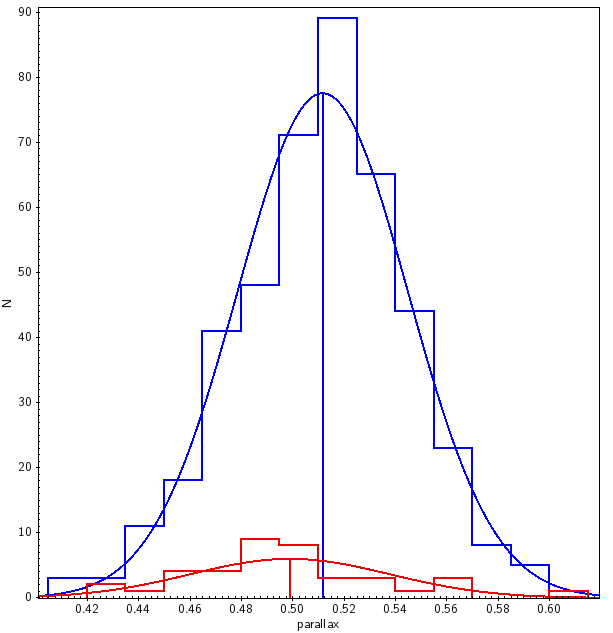}
\caption{The histograms profile for the parallaxes of NGC 6871(blue lines) and Casado 82 (red lines).}
\end{figure}

\begin{figure}
\includegraphics[height=7cm]{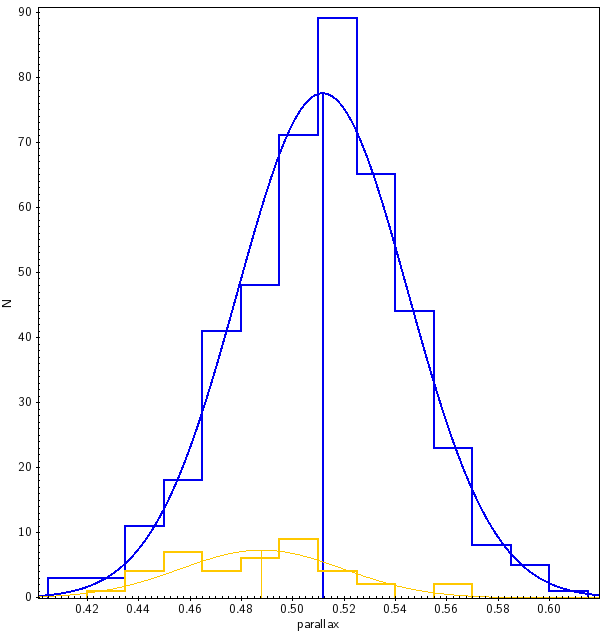}
\caption{The histograms profile for the parallaxes of NGC 6871 (blue lines) and Casado-Hendy 1 (orange
lines).}
\label{figure_13}
\end{figure}

As shown in the precedent sections, all estimated distances indicate that Casado 82 and Casado-
Hendy 1 are slightly farther away than NGC 6871. If, on the one hand, we consider the parallax
distances of NGC 6871 and Casado 82 (Fig. 12), we obtain an average radial distance among them
of 0.1 kpc, with a range of values from 40 to 160 pc from the associated uncertainties. On the other
hand, photometric distances derived for both OCs suggest that their difference in mean distance is
0.13 kpc, widely ranging from 20 to 240 pc due to more significant uncertainties. From the
average values, one wonders if both clusters can be considered members of the same group
(considering the arbitrary threshold distance of 0.1 kpc), even if it is evident from the ensemble of
their properties that the two OCs were born from a unique GMC. Conversely, a minimal radial
distance of 20 pc would figure out a different scenario. The angular separation of the two OCs
centres is 36 arcmin. At an average distance to the system of 1.9 kpc, a projected distance of
almost 20 pc is derived. Combined with the assumed radial distance of 20 pc, the total length
among the two OCs would be 28 pc. Now, if the limiting radius of NGC 6871 is 35 pc (Fig. 3),
Casado 82 would be within the radius of its big sibling, and both clusters would form a binary
system, as suggested in Fig. 11. Nevertheless, the average distances of the two OCs put forward
the opposite: both siblings are gravitationally decoupled and receding from each other at a velocity
of roughly 10 pc/Myr or 10 km/s. Since the projected PMs are very similar for both objects (Fig.
11), we can assume that the substantial velocity difference is in RV. Thus, the RV of Casado 82
should be around 10 km/s more negative (receding) than the RV of NGC 6871. This estimation
agrees with the obtained data: a single star of Casado 82 has an RV of -22 km/s, almost 18 km/s
lower than the median RV of NGC 6871 (Table 2). This difference in RVs also suggests that the
two OCs are diverging from a primordial group in plain disintegration, even if more data on the
RV of Casado 82 are needed to confirm this preliminary conclusion. Similar reasoning suggests an
even more likely dissociation in progress between NGC 6871 and Casado-Hendy 1, taking into
consideration the more considerable projected distance (Figs. 1 and 11), radial distance (Fig. 13),
and an RV difference of ca. 20 km/s (Table 2) among both OCs.\\

As for the mean, minimum and maximum stellar masses, our previous estimates show that they are
all similar for all three OCs, as expected, once more, for OCs born from a single GMC under
similar circumstances.

\section{Concluding remarks}
\label{sec:Concluding remarks}
A primordial group lead by NGC 6871 has been confirmed through Gaia DR3 data and the
existing literature. It is a young complex containing six OCs with ages from 3 to 9 Myr, namely
NGC 6871, Teutsch 8, FSR 198, Biurakan 2, Casado 82 and Casado-Hendy 1. The last two of
these siblings are newly identified and described in detail in the present study. The astrometric and
photometric parameters of the components are sufficiently similar to postulate the case of at least
six clusters born from a single GMC. Three other OCs are suggested as candidates to be members
of the same complex.\\

Despite the significant gravitational pull of NGC 6871, none of the cluster pairs of the group
seems to be gravitationally bound as a binary cluster, with the possible exception of the candidate
pair Teutsch 8/FSR 198 (Song et al. 2022). Instead, NGC 6871 seems to be disintegrating, and the
primordial group members appear to be dispersing out rapidly. Nevertheless, their photometric and
spectroscopic parameters show clearly that all the group members have been born in the same star
formation complex.\\

The present case confirms that the search for new open clusters in the vicinity of young and/or
grouped OCs using Gaia data is an efficient strategy to find new OCs forming primordial groups
(Casado 2021, 2022), and suggests that there are many more OCs still hidden in the plethora of
Gaia results.

\section*{Acknowledgments}
This work made use of data from the European Space Agency (ESA) mission Gaia (https://www.cosmos.esa.int/Gaia, accessed on October 2022), processed by the Gaia Data Processing and Analysis Consortium (DPAC, https://www.cosmos.esa.int/web/Gaia/dpac/consortium, accessed on October 2022). Funding for the DPAC was provided by national institutions, in particular the institutions participating in the Gaia Multilateral Agreement. This research made extensive use of the SIMBAD database and the VizieR catalogue access tool, operating at the CDS, Strasbourg, France (DOI: 10.26093/cds/vizier), and of NASA Astrophysics Data System Bibliographic Services.

\section*{Data availability}
All data used in this study come from the cited astronomical data bases.







%


\end{document}